\documentclass[superscriptaddress,iop,apj]{emulateapj}
\usepackage{latexsym}
\usepackage{amssymb}
\usepackage{amsfonts}
\usepackage{threeparttable}                                                                                           
\usepackage{amsmath}
\usepackage{bm}
\usepackage[normalem]{ulem}
\usepackage{multirow}
\usepackage{natbib} 
\usepackage{subfigure}
\usepackage{color}
\usepackage{calc}
\usepackage{amsmath,amssymb,graphicx}
\usepackage{amssymb,amsmath}
\usepackage{tensor}
\usepackage{bm}
\usepackage{microtype}
\usepackage{booktabs}
\usepackage{times}
\usepackage[varg]{txfonts}
\usepackage[colorlinks, pdfborder={0 0 0}]{hyperref}
\usepackage{subfigure}
\usepackage{lipsum}
\usepackage{breakurl}
\usepackage{graphicx}
\usepackage{pbox}

\newcommand{\igr}{\mbox{IGR J00291+5934}}
\newcommand{\saxj}{\mbox{SAX J1808.4--3658}}

\newcommand{\msun}{\,M_{\odot}}

\newcommand{\ergs}{\rm\,erg\,s^{-1}}

\begin{document}

\title{The Slow Orbital Evolution of the Accreting Millisecond Pulsar IGR J0029+5934}
\shorttitle{11 years of evolution of \igr}
\shortauthors{Patruno et~al.}

\author{
  Alessandro Patruno\altaffilmark{1,2},
\altaffiltext{1}{Leiden Observatory, Leiden University,
              Neils Bohrweg 2, 2333 CA, Leiden, The Netherlands}
   \altaffiltext{2}{ASTRON, the Netherlands Institute for Radio Astronomy, Postbus 2, 7900 AA, Dwingeloo, the Netherlands}}
\begin{abstract}
  The accreting millisecond pulsars \igr\, and \saxj\, are two compact
  binaries with very similar orbital parameters. The latter has been
  observed to evolve on a very short timescale of ${\sim}70$ Myr which
  is more than an order of magnitude shorter than expected.  There is
  an ongoing debate on the possibility that the pulsar spin-down power
  ablates the companion generating large amount of mass-loss in the
  system. It is interesting therefore to study whether \igr\, does
  show a similar behaviour as its twin system \saxj\,. In this work we
  present the first measurement of the orbital period derivative of
  \igr\,. By using \textit{XMM-Newton} data recorded during the 2015
  outburst and adding the previous results of the 2004 and 2008
  outbursts, we are able to measure a 90\% confidence level upper
  limit of
  $-5\times10^{-13}<\dot{P}_b<6\times10^{-13}\rm\,s\,s^{-1}$. This
  implies that the binary is evolving on a timescale longer than
  $0.5\rm\,Gyr$, which is compatible with the expected timescale of
  mass transfer driven by angular momentum loss via gravitational
  radiation. We discuss the scenario in which the power loss from
  magnetic dipole radiation of the neutron star is hitting the
  companion star. If this model is applied to \saxj\, then the
  difference in orbital behavior can be ascribed to a different
  efficiency for the conversion of the spin-down power into energetic
  relativistic pulsar wind and X-ray/gamma-ray radiation for the two
  pulsars, with \igr\, requiring an extraordinarily low efficiency of
  less than $\sim\,5\%$ to explain the observations. Alternatively, the
  donor in \igr\, is weakly/not magnetized which would suppress the possibility
  of generating mass-quadrupole variations.

\end{abstract}



\section{Introduction}

The accreting millisecond X-ray pulsar (AMXP) IGR J00291+5934 is a
peculiar transient X-ray binary for two reasons. The first is that it
is the fastest spinning known accreting pulsar, with a spin frequency
of $\nu\sim599$ Hz. Despite it being still far from the 716 Hz record
holder J1748-2446AD \citep{hes06}, \igr\, has shown a measurable spin-up
during one outburst in 2004 and spin down during quiescence which
allows the determination of its spin evolution over few years
timescale \citep{bur06,pat10b,pap11,har11}.  The second reason, which
is also the main motivation of this work, is that when looking at its
orbital parameters, \igr\, is basically a twin system with the other
well known AMXP SAX J1808.4--3658 (\citealt{wij98,cha98}; see also
Table 1 in \citealt{pat12r} for a comparison). \igr\, was first
discovered during an outburst in December 2004 \citep{sha05,gal05} and
it has been observed in outburst again in
2008~\citep{pat10b,lew10,pap11,har11} and 2015~\citep{san15b}. Its 2008
outburst showed a peculiar behavior, with a first short outburst
lasting $\sim5$ days, followed within one month by a second 12 days
long outburst~\citep{pat10b,har11}.  The source has a radio counterpart
\citep{poo04} and its astrometric position has been determined
with great accuracy~\citep{rup04}.  The binary has an orbital period
of 2.46 hr, so that its orbit is expected to evolve because of angular
momentum loss from gravitational wave emission (see e.g.,
\citealt{pac71,bil01}). Its companion star has a minimum mass of
0.039$\msun$ with a detected optical counterpart both during
outburst~\citep{fox04} and
quiescence~\citep{dav07,tor08,jon08}. In particular, careful
modeling of the optical counterpart in quiescence has shown that the
donor is almost certainly irradiated by a source of energy which is
much more powerful than the quiescent X-ray luminosity available to
the system. The irradiation of the donor from a pulsar wind is a
particularly interesting feature because it has been suggested as the
mechanism at the origin of the peculiar orbital evolution of
\saxj\,~\citep{dis08,bur09} which is expanding on a very short
timescale of $\sim70$ Myr instead of the expected billion years
predicted by the theory of angular momentum loss from gravitational
waves~\citep{har08,pat12}.

In this work we analyze the data from a new set of observations
carried by the \textit{XMM-Newton} observatory during the last
July/August 2015 outburst of \igr\,. We focus on the orbital evolution
of the system since the number of outbursts and the length of the
observational baseline are now sufficient to measure the orbital
period evolution of the binary. Since \igr\, and \saxj\, share so many
similarities it is plausible to expect a similar orbital period
evolution for both systems and in this paper we test this hypothesis. 

\section{Data Analysis}

The 2015 outburst of \igr\, was first detected with the MASTER II
robotic telescope by \citet{lip15} on July 24, 2015 at 05:42:03 UT.
\textit{XMM-Newton} observed the source on July 28, 2015 at 11:48:19
and ended its task on July 29, 2015 at 11:51:02 UT.

We used the European Photon Imaging Camera (EPIC) which
is composed by two MOS CCDs~\citep{tur01} and a pn camera \citep{str01}
sensitive in the 0.1--12 keV range.  In this work we use only the
EPIC-pn data, which are recorded in \emph{TIMING} mode, with
sampling time of about 29.56$\mu$s, sufficient to clearly
detect the accretion powered pulsations.  The data are processed
using SAS version 15.0.0, with the most up-to-date calibration files
(CCF) available on September 2016.

Standard data screening criteria were applied in the extraction of
scientific products with a 0.3--10 keV energy range selected and a net
exposure of 72 ks (after removing solar flares and telemetry
dropouts).  Photons are extracted in a rectangular region
with a width of 6 pixels centered around the RAW coordinate 38 and
only when the PATTERN=0.  The background is obtained from a region of
the same size, at RAWX 2–8. The data are barycentered using the SAS
tool \emph{barycen} by using the source coordinates of \citet{rup04}.

The pulsations are folded in pulse profiles of 32 bins, with a length
of $\sim500$-s each, using a circular Keplerian orbit and a constant
pulse frequency. The first-guess ephemeris are taken from the 2004
outburst (see e.g., \citealt{pat10b}) with the time of passage to the
ascending node ($T_{\rm asc}$) updated from \citet{kui15}, which
performed a first timing analysis of the 2015 outburst with
\textit{INTEGRAL} data.

\section{Results}

Since the pulse profiles of \igr\, are nearly sinusoidal, we define the pulse
time of arrivals (ToAs) as the peak of the sinusoid of each profile.
We then fit the ToAs with the software TEMPO2
(v. 2016.05.0; \citealt{hob06}) by using a constant pulse frequency plus a
constant circular Keplerian orbit (ELL1 model). We then refine the
ephemeris by iterating the procedure until convergence is achieved.
We refer to the pulse frequency (observable) as distinct from the spin
frequency since it has been shown that
the X-ray flux has an influence of the pulse ToAs and might affect the
determination of the correct spin frequency up to several tenths of
$\mu$Hz \citep{har08,pat10b,pat09f}.  We find a pulse frequency of
$\nu=598.89213099(6)$ Hz and no pulse frequency derivative is detected
with $|\dot{\nu}|<10^{-11}\rm\,Hz\,s^{-1}$ at the 95\% confidence
level.

To detect the evolution of the orbit we instead follow the procedure already outlined in \citet{pat12}, which is also used in \citet{har08,dis08,bur09,har09,bur10,san16}, i.e., we select all four measured $T_{\rm asc}$ from the 2004, the double 2008 and the 2015 outbursts and we use
the quantity $\Delta\,T_{\rm asc}=T_{\rm asc,i}-\left(T_{\rm asc,ref}+N\,P_{b}\right)$, where
$T_{\rm asc,i}$ refers to the $i-th$ outburst, $N$ is the closest integer to
$\left(T_{\rm asc,i}-T_{\rm asc,ref}\right)/P_b$ and $P_b$ is the orbital period.
Since the best determination of $P_b$ is made in 2004, we use that outburst as the
reference one in our first set of calculations (see e.g. Table~3 in \citealt{pat10b}). 
We use a polynomial expansion to describe the evolution of the
time of passage through the ascending node:
\begin{equation}
T_{\rm asc}(N) = T_{\rm asc,ref}+P_bN+\frac{1}{2}P_b\dot{P}_bN^2+...
\end{equation}

\begin{figure}[t]
  \begin{center}
    \rotatebox{-90}{\includegraphics[width=0.75\columnwidth]{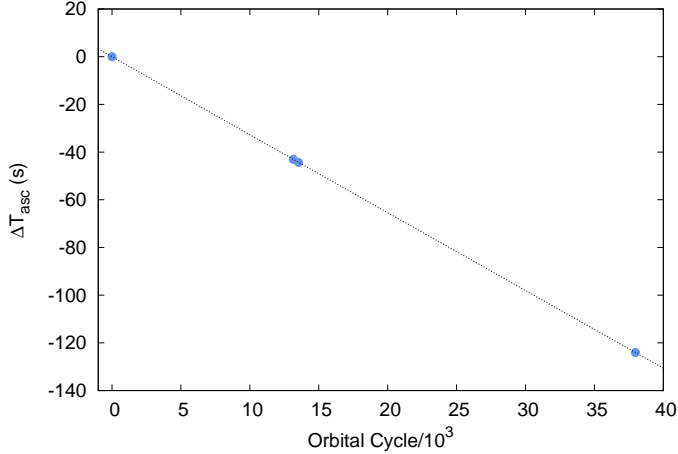}}
  \end{center}
  \caption{The differential corrections to the time of passage through
    the ascending node found when using the value of $T_{\rm asc}$ as
    reported in \citet{pat10b}. A linear trend (dotted line) is visible and can be
    well fitted by shifting the orbital period by ${\sim}3.3$ ms. The error bars
    of the data are smaller than the symbols used.
 }\label{fig1}
\end{figure}

In our analysis we calculate first the differential correction to the
orbital period $\delta\,P_b$ by fitting $\Delta\,T_{\rm\,asc}$ with a
linear function $\Delta\,T_{\rm asc} = \delta\,P_b\,N$. The fit gives
$\delta\,P_b=3.266(2)$ ms with a $\chi^2/dof=0.15/3$ (see
Figure~\ref{fig1}). The origin of the very small $\chi^2$
indicates that the statistical errors on the fitted parameters are
unrealistic. The 2004 and 2015 outburst have errors on
$T_{\rm asc}$ which are a factor of 5--10 smaller than the two 2008
outbursts.  Therefore when fitting a linear function there is little
contribution from the 2008 data points and the fit gives a very small
$\chi^2$.

The linear trend is very evident, so we use the best-fit $\delta\,P_b$
to correct the orbital period. We then re-analyze the data published
in \citet{pat10b} for the entire data-set recorded for \igr\, by
folding the 2004, 2008 and 2015 data with the new orbital period and
fitting the ToAs of each of the four outbursts with a Keplerian orbit
where $P_b$ is now fixed as well as the projected semi-major axis of
the orbit and we fit only $T_{\rm asc}$. This gives a new set of
improved $T_{\rm asc}$ which we report in Table~\ref{tab:tasc}. We
also tried to detect variations of the projected semi-major axes of
the orbit $a_1$. The four $a_1$ show no trend and are well fit by a
constant ($\chi^2/dof=4.2/3$).
\begin{table}                                                                             
\centering                                                                                
\caption{Time of Passage through the Ascending Node for \igr\,}                                   
\begin{tabular}{llll}                                                                     
\hline                                                                                    
\hline                                                                                    
Outburst & $T_{\rm asc}$ [MJD]& Stat. Error [MJD]\\                                           
\hline
\smallskip
2004 & 53345.1619259 & 0.0000016\\
\smallskip
2008 (1st) & 54692.0411119 & 0.0000018\\
\smallskip
2008 (2nd) & 54730.5292226 & 0.0000015\\
\smallskip
2015 & 57231.8470383 & 0.0000006\\
\hline                                                                                    
\end{tabular}\label{tab:tasc} 
\end{table}

\begin{figure}[t]
  \begin{center}
    \rotatebox{-90}{\includegraphics[width=0.75\columnwidth]{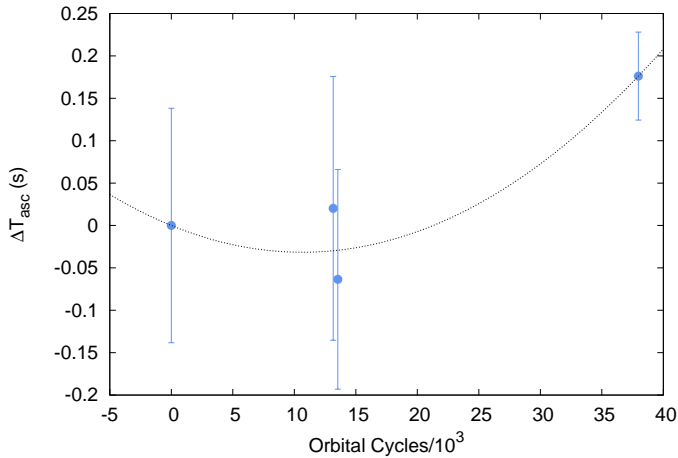}}
  \end{center}
  \caption{The differential corrections to the time of passage through
    the ascending node after correcting $P_b$. A constant function
    fits well the data and a quadratic polynomial (dotted line)
    provides upper limits on the presence of an orbital period
    derivative $-5\times10^{-13}<\dot{P}_b<6\times\,10^{-13}$
    (90\% confidence interval). }\label{fig2}
\end{figure}

We then inspect the new $\Delta\,T_{\rm asc}$ to see whether residual
trends are observed. For example, in \saxj\, a clear polynomial trend
is observed \citep{pat12,har08,dis08} which is interpreted as an
expansion of the orbit.  In \igr\, the residuals show instead very
little structure, which is indicative of a very slow variation of the
orbit. The data can indeed be well fitted with a constant consistent
with zero.  A fit with a quadratic polynomial gives both the linear
and quadratic term consistent with zero (see Fig.~\ref{fig2}) with a
$\chi^2=0.18$ for 2 degrees of freedom. We therefore can set a 90\%
confidence interval for any orbital period derivative of
$-5\times10^{-13}<\dot{P}_b<6\times10^{-13}$. This means that the
orbital evolution timescale of \igr\, is at least
$\tau>\frac{P_b}{\dot{P}_b}\sim0.5$ Gyr. The final orbital ephemeris
of \igr\, are reported in Table~\ref{tab:sol}.

\begin{table}                                                                             
\centering                                                                                
\caption{\igr\, Orbital Solution}                                   
\begin{tabular}{llll}                                                                     
\hline                                                                                    
\hline                                                                                    
Parameter & Value & Stat. Error\\                                           
\hline                                                                                    
$T_{\rm asc}$ [MJD] & 57231.8470383 & $6\times10^{-7}$ \\                             
$P_b$ [s]    & 8844.07673 & $9\times10^{-5}$ \\
$\dot{P}_b$  [$10^{-13}\rm\,s/s$] & (-5; 6) & (90\% c.l.)\\ 
$a_1$ [lt-ms] & 64.993 & 0.002  \\
$e$ &  $<0.0002$ &  (95\% c.l.)  \\
$P_{epoch}$ [MJD] &  57300&\\
\hline                                                                                    
\end{tabular}\label{tab:sol} 
\end{table}

As a further test we also combined all the data from 2004 up to 2015
in a single sequence of ToAs and fitted a Keplerian
orbital solution with a $\dot{P}_b$ term with TEMPO2.  The orbit can
be phase-connected because the total number of cycles observed is
$N_{cycles}{\sim}40,000$ and the initial error on our orbital period
is $\sigma_{P_{b}}=0.002$ s (the 2004 orbital period, see
\citealt{pat10b}) so that $\sigma_{P_b}\lesssim P_b/N_{\rm
  cycles}\approx 0.2\rm\,s$. We stress that the pulse frequency (and
its first time derivative) are very weakly covariant with the
Keplerian parameters, so that any unmodeled trend in the neutron star
spin is not affecting the determination of the orbital
solution.  The results are fully compatible within one sigma with
those reported in Table~\ref{tab:sol} and give also compatible statistical
uncertainties and confidence intervals.

\section{Discussion}

The orbital evolution of \igr\, proceeds on a timescale $>500\rm\,Myr$ which
is in line with the expectation of a binary evolving via angular
momentum loss caused by gravitational wave radiation.  Indeed in this
case the evolutionary timescale is~\citep{pac71}:
\begin{equation}\label{eq:gw}
\tau_{\rm\,gw} = 380 \frac{(1+q)^{2}}{q}\left(\frac{M_1+M_2}{\msun}\right)^{-5/3}\left(\frac{P_b}{\rm days}\right)^{8/3}\rm\,Gyr.
\end{equation}
where $q=M2/M1$ is the binary mass ratio and $M_1$ an $M_2$ are the
neutron star and donor mass. If we use reasonable values of
$M_1=1.4\msun$, $M_2=0.1\msun$ then $q\simeq0.07$ and $\tau_{\rm
  gw}\simeq\,7$ Gyr, consistent with the upper limits observed in this
work. Although there is still the possibility that \igr\, evolves on a
timescale shorter than predicted (a factor $\sim10$ is still allowed
by our upper limits on $\dot{P}_b$), the binary seems to pose at the
moment no challenge when compared to the expected behavior of binary
evolution. However, when compared to the behavior of the AMXP \saxj\,
the results presented in this paper become difficult to interpret.
Indeed the ``twin'' system SAX J1808.4--3658 has shown an expansion of
the orbit with a $\dot{P}_b\sim3.5\times10^{-12}$, which is about one
order or magnitude larger than our upper limits on \igr\,. The
interpretation on why the orbits of these two systems behave so
differently is puzzling if we look at all other measured parameters of
the two binaries. \saxj\, shows indeed very similar properties: its
orbit is 2.01 hours, its minimum companion star is
0.043$\msun$~\citep{cha98} and the companion is irradiated by a
powerful source of energy~\citep{hom01,del08,wan13}. It has been
proposed that, for both AMXPs, the source of extra irradiation comes
from the spin-down power of the neutron star which might turn on
during quiescence and irradiate with its powerful wind the exposed
face of the donor star~\citep{bur03,cam04,dav07}.  To understand
whether such scenario is energetically feasible for \igr\, we can
assume two extreme cases, one with the minimum donor mass and with a
neutron star mass of $M_1=2.5\msun$ and the second with the maximum
donor mass and $M_1=1.2\msun$. With these parameters we can calculate
the orbital separation and the donor Roche lobe radius for the two
most extreme mass ratios.  The fraction of intercepted pulsar
radiation/wind can be estimated with the simple expression
$f=(R_{2}/2A)^2$, where $R_2$ is the Roche lobe radius and $A$ is the
orbital separation. This fraction is nearly identical between \igr\,
and \saxj\,, $f=0.4$--$1.0\%$, i.e., their donors are absorbing equal
fractions of input energy.

Beside the excessive optical luminosity of the companion, there is
further evidence (in both systems) that the neutron star is indeed
losing power most likely due to magnetic-dipole radiation during
quiescence. \igr\, has been observed to spin-down in quiescence at a
rate of $3-4\times10^{-15}\rm\,Hz\,s^{-1}$~\citep{pat10b,pap11,har11}
and SAX J1808.4--3658 is seen to spin down at a similar rate of
${\sim}10^{-15}\rm\,Hz\,s^{-1}$~\citep{har08,har09,bur09,pat12}. The
assumption that the source of spin-down is the magnetic dipole
radiation has lead to the indirect measurement of the neutron star
magnetic field in both systems: $1.5-2.0\times10^{8}$ G for \igr\, and
$1$--$3\times10^{8}$ G in SAX J1808.4--3658~\citep{har08,pat12,pap09}.
The spin-down power available in the two systems is therefore of the
same order of magnitude, and it is within a factor 5 if we assume that
the moment of inertia is the same for the two neutron stars.  Other
similarities between the two binaries include the observation of
thermonuclear X-ray bursts \citep{boz15,int98}, the presence of
$\rm\,H\alpha$ emission line in outburst~(\citealt{roe04}, Kaper
private communication) the detection of transient radio signals during
outbursts~\citep{poo04,gae99}, a similar recurrence time for the
outbursts (3--4 years for \saxj\, and 4--6 years for \igr\,) and
comparable mass transfer and accretion rates~\citep{bil01,gal05}.  It
is safe to say that, with the exception of the orbital evolution,
\igr\, and \saxj\, have a tight match between all other observable
quantities.  It follows that if \textit{any} of the observables listed
above (see a summary in Table~\ref{tab:summary}) is used to support a specific
interpretation of the fast orbital evolution of \saxj\,, the same should be
true for \igr\,.

For example, the aforementioned over-luminous optical counterparts of
\igr\, and \saxj\, in quiescence have been interpreted as being
generated by irradiation of the donor from the pulsar
radiation/wind. In particular, the power injected by the pulsar into
the companion of \saxj\, has been suggested to generate a large
mass-loss~\citep{dis08,bur09} which in turn would explain the large
orbital $\dot{P}_b$. In this highly non conservative mass-transfer
scenario about $99\%$ of the mass transferred is lost in a stellar
wind.  \igr\, has an over luminous optical companion, and this has
been interpreted too as due to irradiation from the pulsar wind. A
similar interpretation for \igr\, seems difficult to reconcile with
its slow orbital evolution that at the moment is compatible with a
conservative scenario (i.e., no mass-loss).  Indeed the variation of
the orbital period of the binary as a consequence of a spherical wind
loss from the donor is~\citep{fkr02}:
\begin{equation}\label{ml}                                                                                            
\frac{\dot{P}_b}{P_b} = -2\frac{\dot{M}_2}{M_2}                                                                       
\end{equation} 
If we use our upper limit on $\dot{P}_b$, then: 
\begin{equation}
\dot{M}_c = \frac{1}{2}\frac{\dot{P}_b}{P_b}\,M_c\lesssim 10^{-10}\msun\,yr^{-1}
\end{equation}
which is about an order of magnitude smaller than proposed for example
in \saxj\,. To explain such dramatic difference in behavior between
these two accreting systems we propose three possibilities.

\begin{table}
  \begin{threeparttable}
\centering                                                                                
\caption{Comparison between observables in \igr\, and \saxj\,}
\begin{tabular}{lll}                                                                     
\hline                                                                                    
\hline                                                                                    
Parameter & \igr\, & \saxj\,\\
\hline
\smallskip
Min. Donor Mass [$\msun$] & 0.039 & 0.043\\
Max. Donor Mass$^{A}$ [$\msun$] & 0.09 & 0.10 \\
Donor Radius [$R_{\odot}$] & 0.13--0.20 & 0.11--0.17\\
Orbital Period [hr] & 2.46 & 2.01 \\
Proj. Semi-major axis [lt-ms] & 64.993 & 62.812\\
Outb. Recurrence Time [yr] & 4--6 & 3--4\\
Irradiation$^{B}$ [$10^{33}\ergs$]& ${\sim}4-8$  & ${\sim}1-10$\\
$L_{\rm sd}$$^{C}$ [$10^{34}\ergs$] & 7 & 2\\
Intercepted power $f$& (0.4--1.0)\% & (0.4--1.0)\%\\
\hline                                                                                    
\end{tabular}\label{tab:summary} 
\begin{tablenotes}
      \small
    \item $^A$ the maximum companion mass is a $90\%$ confidence level upper bound assuming $i=26^{\circ}$\citep{hob06}.
      \smallskip
    \item $^B$ this parameter indicates the minimum power required to produce sufficient irradiation  to explain the optical counterpart. For \saxj\, the irradiation luminosity is given for a range of distances 2.5--3.5~kpc
      \smallskip
      \item$^{C}$ Spin down luminosity available in the system via $I\omega\dot{\omega}$.
      \item The table summarizes the properties of \igr\, and \saxj\, relevant for the orbital evolution of the systems (see main text for an explanation and references).
\end{tablenotes}
\end{threeparttable}
\end{table}

A first possibility is that there are two different mechanisms
operating in these binaries. There is a subtle difference in the
optical behavior of \igr\, with respect to \saxj\, during quiescence
that was reported by \citet{jon08}. If the optical excess can be
entirely ascribed to the donor being irradiated then, during
quiescence, an optical modulation with a peak at the neutron star
inferior conjunction (phase 0.5) should be observed.  This was indeed
the case for \igr\, as reported by \citet{dav07} when using
near-infrared observations. However, \citet{jon08} when analyzing
further observations in the $I$-band (September 13 and 14 2006) found
a sinusoidal modulation peaking at phase $0.34\pm0.03$ along with very
large (${\sim}1$ mag) optical flares. The conclusion of \citet{jon08}
was that a modulation with a period slightly different from the
orbital period might be responsible for the sinusoidal modulation
observed, in a way similar to superhumps observed in cataclysmic variables
and other X-ray binaries. However, the (quasi)orbital modulation
detected by \citet{jon08} and \citet{dav07} has an amplitude of only a
few percent, meaning that the bulk excess optical light still exceeds
by 1--2 orders of magnitude the quiescent X-ray luminosity of the
residual accretion disk. Given that any other observable quantity is
almost identical in both AMXPs, we cannot support any other
possibility with the existing observations.

The second possibility is that the mass-loss scenario in \saxj\, is
not correct and that the orbital period derivative is caused by a
different phenomenon. \citet{har08, har09} and \citet{pat12} proposed
a scenario in which quadrupolar mass variations in the donor cause a
variation of the orbital parameters due to spin-orbit
coupling~\citep{app92, app94}.  In that scenario the donor is
required to have a large magnetic field for the effect to take place,
and in \saxj\, it was estimated that a field of the order of $1$~kG is
necessary.  The Applegate mechanism (or a similar one) is appealing
because it might explain the observation in terms of an unseen
magnetic field of the donor star (which, in the case of \igr\, should
be weakly or no magnetized). However, it is not clear whether such a
mechanism can take place in a tiny donor star like those observed here
and indeed several criticisms exist in the literature.  The main
objection to the model is that the formation of a mass quadrupole
requires a certain amount of energy that is too large when compared to
the nuclear energy budget of the donor or to the tidal dissipation in the
system (see e.g. \citealt{bri06}).

Given that the fraction of absorbed power $f$ is identical in both
AMXPs, the final possibility is that the irradiation of the donor
proceeds in the two systems with different efficiency. For example,
\citet{dav07} estimated that the power required to irradiate the donor
of \igr\, is $4\times10^{33}\ergs$ (\saxj\, requires a similar value,
see \citealt{bur03,cam04}), whereas the spin-down power available in
the system is $L_{sd}=I\omega\dot{\omega}\approx8\times10^{34}\ergs$,
where $I=10^{45}\rm\,g\,cm^2$ is the moment of inertia of the neutron
star, and $\omega$ and $\dot{\omega}$ are the angular frequency and
its first time derivative (which come from observations).  This
requires that \textit{less} than ${\sim}5\%$ of the spin-down power is
converted into energetic wind and X-ray/gamma-ray radiation. For
\saxj\, instead, the spin down-power is $2\times10^{34}\ergs$ and
therefore the efficiency required is of the order of $40\%$ with a
peak efficiency close to 100\% (Patruno et al. 2016).

\section{Conclusion}

We have placed stringent constraints on the orbital evolution of the
accreting millisecond pulsar \igr\,. We find an upper limit for the
orbital period derivative which translates into an orbital evolution
timescale larger than 0.5 Gyr. There is a substantial difference
between this behavior and that of \saxj\,, an AMXP with very similar
orbital parameters and donor properties. We find that, if we want to
explain the orbital evolution of both binaries with a mass-loss model
due to irradiation of the companion, then the pulsar in \igr\, is
radiating power which is partially converted into winds and high
energy photons with an efficiency of less than 5\%. Alternatively, if
the variations of the orbit seen in \saxj\, are due to spin-orbit
coupling, then the donor star in \igr\, should be weakly or no
magnetized which would suppress the mass-quadrupole variations.

\bigskip

\acknowledgements{I acknowledge support
  from an NWO Vidi fellowship.}


\end{document}